\documentclass{PoS}

\title{The Hadronic Spectrum and Confined Phase in (1+1)-Dimensional Massive Yang-Mills Theory}

\ShortTitle{The Hadronic Spectrum and Confined Phase in (1+1)-Dimensional Massive Yang-Mills Theory}

\author{\speaker{Axel Cort\'es Cubero}\thanks{This talk includes work done in collaboration with Peter Orland}\\
	Baruch College, The City University of New York, 17 Lexington Avenue, New York, NY 10010, U.S.A. ,\\
	\\
        The Graduate School and University Center, The City University of New York, 365 Fifth Avenue, New York, NY 10016, U.S.A. and\\
        \\
        Scuola Internazionale Superiore di Studi Avanzati, via Bonomea 265, 34136 Trieste, Italy.\\

        E-mail: \email{acortes\_cubero@gc.cuny.edu}}

\abstract{Massive Yang-Mills theory is known to be renormalizable in 1+1 dimensions. The gluon mass is introduced by coupling the gauge field to an SU(N) principal chiral nonlinear sigma model. The proof of renormalizability relies on the asymptotic freedom of the sigma model. However, renormalization forces the gluon mass to infinity. The continuum theory is in a confined phase rather than a Higgs phase. The physical excitations of the system are hadron-like bound states of sigma model particles. We calculate the massive spectrum of meson-like bound states analytically, using the exact S-matrix of the sigma model. The baryon-like spectrum can be found in principle by solving a quantum mechanical N-body problem. We remark on the evidence for the confined phase found for SU(2) in recent lattice simulations by Gongyo and Zwanziger. Their simulations show evidence for a Higgs-like phase which seems to disappear with increasing volume, finding agreement with our analysis in the continuum.}

\FullConference{The 32nd International Symposium on Lattice Field Theory,\\
		23-28 June, 2014\\
		Columbia University New York, NY}

\begin{document}

\section{Introduction}
In this talk we discuss the spectrum of Massive Yang-Mills theory in 1+1 dimensions. This model is renormalizable, in contrast to the same model in higher dimensions \cite{renormalizable}. Renormalizability is proven by noticing that the Massive Yang-Mills action is equivalent to a gauged principal chiral sigma model (PCSM), which is asymptotically free.

The action of the PCSM is
\begin{eqnarray}
S=\frac{N}{2g^2}\int d^2x\,{\rm Tr}\partial_\mu U^\dag(x)\partial^\mu U(x),\label{pcsm}
\end{eqnarray}
with $U\in SU(N)$. We review key aspects of the PCSM, including its integrability, in the next section. The action (\ref{pcsm}) has an $SU(N)\times SU(N)$ global symmetry given by $U(x)\to V^L U(x) V^R$, with $V^{L,R}\in SU(N)$. We call the Noether currents associated with these symmetries, $J^L(x)$, and $J^{R}(x)$.

We promote one of the $SU(N)$ global symmetries (the left handed $V^L$) of \label{pcsm} to a global gauge symmetry. The action of the gauged sigma model is 
\begin{eqnarray}
S=\int d^2x -\frac{1}{4}{\rm Tr} F_{\mu\nu} F^{\mu\nu} +\frac{1}{2g_0^2} {\rm Tr}D_\mu U^{\dag} D^\mu U,\label{gaugedpcsm}
\end{eqnarray}
with $D_\mu=\partial_\mu+i e A_\mu^{L}$.
It can be seen that the action (\ref{gaugedpcsm}) is that of massive Yang-Mills theory by looking at the unitary gauge $U(x)=1$, where
\begin{eqnarray}
S=\int d^2x -\frac{1}{4}{\rm Tr}F_{\mu\nu}F^{\mu\nu} -\frac{e^2}{2g_0^2}{\rm Tr} A_\mu A^\mu.\label{unitarygauge}
\end{eqnarray}

From looking at the action (\ref{unitarygauge}) one would naively guess that the particles of this theory are gluons with mass $e/g_0$. However, the asymptotic freedom of the PCSM forces the bare coupling $g_0$ to vanish. The gluon mass then diverges, and one cannot observe this gluon at low energies.

The spectrum of massive Yang-Mills theory in two dimensions does not consist of massive gluons. The question we address in this talk is, what then are the particles of this model. 

The main point we present is that two-dimensional massive Yang-Mills theory is not in a Higgs-like phase, but in a confined phase.  The physical particles of this model are not massive gluons, but hadron-like bound states of sigma model particles. This is different than 2+1 and 3+1 dimensions, where both phases are present \cite{phases}.

\section{Review of the Principal Chiral Sigma Model}

In this section, we recall previous exact results that have been obtained using the integrability of the PCSM. Integrability in quantum field theories implies that all scattering events are elastic and factorizable. There is no particle creation, the set of particle momenta is conserved, and any scattering can be written as the product of two-particle S-matrices. These properties have been used extensively to calculate exact S-matrices \cite{zamolodchikov}. In particular, the S-matrix of two PCSM particles is known \cite{wiegmann}:
\begin{eqnarray}
&&\left._{\rm out}\langle P,\theta_1',c_1,d_1;P,\theta_2',c_2,d_2|P,\theta_1,a_1,b_1;P,\theta_2,a_2,b_2\rangle\right._{\rm in}=S(\theta,N)^{c_1, d_1; c_2, d_2}_{a_2, b_2; a_1, b_1}\langle \theta_1^\prime\vert \theta_1\rangle \langle \theta_2^\prime\vert \theta_2\rangle \nonumber\\
&&\,\,\,\,\,\,\,\,\,\,\,\,\,\,\,\,\,\,\,\,\,=S(\theta,N)\left(\delta_{a_1}^{c_1}\delta_{a_2}^{c_2}-\frac{2\pi i}{N\theta}\delta_{a_1}^{c_2}\delta_{a_2}^{c_1}\right)\times\left(\delta_{b_1}^{d_1}\delta_{b_2}^{d_2}-\frac{2\pi i}{N\theta}\delta_{b_1}^{d_2}\delta_{b_2}^{d_2}\right)\langle \theta_1'|\theta_1\rangle\langle \theta_2'|\theta_2\rangle,\label{smatrix}
\end{eqnarray}
where $P$ labels a particle, $A$ labels an antiparticle and $\theta_i$ is the rapidity of the $i$-th particle, defined by the parametrization of energy and momentum: $E_i=m\cosh\theta_i,\,\,p_i=m\sinh\theta_i$, and $\theta=\theta_1-\theta_2$.  The $i$-th particle has a left-color index $a_i$ and a right color index $b_i$, with
\begin{eqnarray}
S(\theta)=\frac{\sinh(\frac{\theta}{2}-\frac{\pi i}{N})}{\sinh(\frac{\theta}{2}+\frac{\pi i}{N})}\left[\frac{\Gamma(i\theta/2\pi +1)\Gamma(-i\theta/2\pi-\frac{1}{N})}{\Gamma(i\theta/2\pi+1-\frac{1}{N})\Gamma(-i\theta/2\pi)}\right]^2. \nonumber
\end{eqnarray}
The particle-antiparticle S-matrix can be found by crossing symmetry, $\theta\to\pi i-\theta$.

Knowledge of the exact S-matrix is the starting point of the integrable bootstrap program. One can use the S-matrix to calculate exact form factors (matrix elements of local operators). Form factors of the PCSM in the 'tHooft large-$N$ limit for different operators have been found in References \cite{orland}, \cite{multiparticle}, \cite{correlation}. At finite $N$, only the first nontrivial form factors have been found (with only two-particle states). Of particular interest to us is the form factor of the Noether current operator \cite{multiparticle}
\begin{eqnarray}
\langle 0 \vert \!\!\!&&\!\!\!j_\mu^L(0)_{a_0c_0}\vert A,\theta_1,b_1,a_1;P,\theta_2,a_2,b_2\rangle=(p_1-p_2)_\mu\left( \delta_{a_0a_2}\delta_{c_0a_1}-\frac{1}{N}\delta_{a_0c_0}\delta_{a_1a_2}\delta_{b_1b_2}\right)
\nonumber\\
&&\times\frac{2\pi i}{(\theta+\pi i)}\exp \int_0^\infty \frac{dx}{x}\left[\frac{-2\sinh\left(\frac{2x}{N}\right)}{\sinh x}+\frac{4e^{-x}\left(e^{2x/N}-1\right)}{1-e^{-2x}}\right]\frac{\sin^2[x(\pi i-\theta)/2\pi]}{\sinh x}.\label{formfactor}
\end{eqnarray}
for $N>2$. The form factors in the case $N=2$ are equivalent to the form factors of the isovector-valued $O(4)$ nonlinear sigma model, by virtue of $SU(2)\times SU(2)\simeq O(4)$. The two-particle form factors of the $O(N)$ sigma model have been found in Ref. \cite{karowskiweisz}.

\section{The Bound-state Spectrum in Massive Yang-Mills}
In this section, we will omit writing color indices, for simplicity and clarity. The Hamiltonian of massive Yang-Mills theory in the completely-fixed axial gauge, $A_0=0,\,A_1(t=0)=0$, is found in detail in Ref. \cite{dynamicalmass}, to be
\begin{eqnarray}
H=H_{\rm PCSM} -\frac{e^2}{2g_0^4}\int dx^1 \int dy^1\vert x^1-y^1\vert j_0^L(x^1)j_0^L(y^1).\label{axialhamiltonian}
\end{eqnarray}
The Hamiltonian (\ref{axialhamiltonian}) appears nonlocal, but this is a natural consequence of the axial gauge. It can be made local again by re-introducing the temporal component of the field $A_0$.

The Hamiltonian (\ref{axialhamiltonian}) describes sigma-model particles confined by a linear potential with string tension $\sigma=e^2C_N$, where $C_N$ is the smallest eigenvalue of the Casimir operator of $SU(N)$. This can be seen by interpreting the temporal component of the current $J_\mu^L$ as the left-color charge density. The physical eigenstates of (\ref{axialhamiltonian}) are meson-like particle-antiparticle bound states of mass $M=2m+E$, or baryon-like $N$-particle bound states. One can in principle find the bound-state spectrum by calculating the wave function and  eigenvalues of (\ref{axialhamiltonian}). We have found this wave function for the meson-like bound states in the nonrelativistic limit, in Reference \cite{dynamicalmass}. A relativistic approach for treating confinement in integrable field theories with nonintegrable deformations has been introduced in \cite{ffpt}, which goes beyond the level of our analysis of this model.

We now present our results for the nonrelativistic limit where $m\gg e$. The meson-like state consists of a particle at position $x^1$, and an antiparticle at $y^1$. Using center of mass coordinates $x=x^1-y^1$, the meson wave function, $\Psi(x)$, satisfies the nonrelativistic Schrödinger equation
\begin{eqnarray}
-\frac{1}{m}\frac{d^2}{dx^2}\Psi(x)+\sigma \left\vert x \right\vert \,\Psi(x)=E\Psi(x),\label{schroedinger}
\end{eqnarray}
The solution to Eq. (\ref{schroedinger}) is
\begin{eqnarray}
\Psi(x)=
C Ai\left[(m\sigma)^{\frac{1}{3}}\left(\vert x\vert-\frac{E}{\sigma}\right)\right],\label{airy}
\end{eqnarray}
where $Ai(x)$ is an Airy function of the first kind, and $C$ is a normalization constant.

A quantization condition for the binding energy $E$ arises from the requirement that the wave function (\ref{airy}) becomes the wave function of a free particle antiparticle pair as $x\to0$ (because the linear potential vanishes). The wave function of a free particle antiparticle pair is
\begin{eqnarray}
\Psi(x^1,y^1)=\left\{\begin{array}{c}
e^{ip_1x^1+ip_2y^1},\,\,\,\,\,\,\,\,\,\,\,\,\,{\rm for}\,\,x^1<y^1\\
\,\\
e^{ip_2x^1+ip_1y^1}S(\theta),\,\,\,\,\,\,{\rm for}\,\,x^1>y^1\end{array}\right.,\label{freewave}
\end{eqnarray}
where $S(\theta)$ is the particle-antiparticle S-matrix.

By making (\ref{airy}) and (\ref{freewave}) be equivalent as $x\to0$, we find the quantized energy spectrum
\begin{eqnarray}
E_n=\left\{\left[\epsilon_n+\left(\epsilon_n^2+\beta_N^3\right)^{\frac{1}{2}}\right]^{\frac{1}{3}}+\left[\epsilon_n-\left(\epsilon_n^2+\beta_N^3\right)^{\frac{1}{2}}\right]^{\frac{1}{3}}\right\}^{\frac{1}{2}},\nonumber 
\end{eqnarray}
where
\begin{eqnarray}
\epsilon_n=\frac{3\pi}{4}\left(\frac{\sigma}{m}\right)^{\frac{1}{2}}\left(n+\frac{1}{2}\pm \frac{1}{4}\right),\,\,\,\,\,\,\,\,\,\,\,\,\,\nonumber
\end{eqnarray}
\begin{eqnarray}
\beta_N=\frac{\sigma^{\frac{1}{2}}}{2\pi m}\int_0^\infty \frac{d\xi}{\sinh \xi}\left[2(e^{2\xi/N}-1)-\sinh(2\xi/N)\right], \nonumber
\end{eqnarray}
The meson states have masses $M_n=2m+E_n$.

The baryon mass spectrum can be found in principle by the same method. The only difficulty is that one has to solve a $N$-body Schrödinger equation with  potential
\begin{eqnarray}
V(x_1,\dots, x_N)=\sum_{i=1}^{N-1}\sum_{j=i+1}^{N} \sigma \vert x_i^1-x_j^1\vert.\nonumber
\end{eqnarray}
The $N$-body wave function is much harder to find exactly, and numerical methods are necessary. Recently, there have been some efforts in this direction for the related problem of the baryon spectrum in the three-states Potts model in an external magnetic field \cite{potts}.

\section{Correlation Functions}
With knowledge of the bound-state wave function (\ref{airy}) and the two-particle form factor (\ref{formfactor}) we can find approximate correlation functions which work well at large distances. The first step is to find the form factor of an operator $\mathcal{A}(x)$ with a bound state of mass $M_n$:
\begin{eqnarray}
\langle 0\vert\mathcal{A}(x)\vert B, \phi, n\rangle,\label{mesonformfactor}
\end{eqnarray}
where $B$ denotes that the excitation is a meson bound state, $\phi$ is the meson's rapidity, and $n$ is the meson's energy level.

In the nonrelativistic limit, we can evaluate (\ref{mesonformfactor}) using the "two-quark" approximation originally presented in Ref. \cite{zamolodchikovfonseca},
\begin{eqnarray}
\vert B, \phi, n\rangle\approx e^{ix^1M_n \sinh\phi}\frac{1}{\sqrt{m}}\int_{-\infty}^\infty\frac{d\theta}{4\pi} \Psi_n(\theta)\,\vert A,\theta,a_1,b_1;P,-\theta,a_1,b_1\rangle,\label{twoquark}
\end{eqnarray}
where $\Psi_n(x)$ is given by substituting $E_n$ in Eq. (\ref{airy}), and $\Psi_n(\theta)$ is its Fourier transform. There are relativistic corrections to (\ref{twoquark}) from states with a higher number of particles which we ignore. The one-meson form factor of  an operator $\mathcal{A}$ is written in terms of the two-sigma-model-particle form factor:
\begin{eqnarray}
\langle 0\vert \mathcal{A}(x)\vert B,\phi,n\rangle&=&e^{s\phi}e^{ix^1M_n \sinh\phi}\int dz\int\frac{d\theta}{4\pi} e^{izm\sinh\theta}\frac{1}{\sqrt{m}}\left(\frac{E_n}{\sigma^H}\right)^{\frac{1}{4}} {\rm Ai}\left[(m\sigma^H)^{\frac{1}{3}}\left(\vert z\vert -\frac{E_n}{\sigma^H}\right)\right]\nonumber\\
&&\times\langle 0\vert \mathcal{A}(x)\vert A,\theta,a_1,b_1;P,-\theta,a_1,b_1\rangle\nonumber
\end{eqnarray}
We find a two-point correlation function, $D^{\mathcal{A}}(x)=\langle0\vert \mathcal{A}(x)\mathcal{A}(0)\vert0\rangle$, by summing over intermediate particle states:
\begin{eqnarray}
&&\mathcal{D}^{\mathcal{A}}(x)=\langle 0\vert \mathcal{A}(x)\vert 0\rangle \langle 0\vert \mathcal{A}(0)\vert 0\rangle\nonumber\\
&&\,\,+\sum_{n=1}^{n_s}\int\frac{d\phi}{4\pi}e^{-ix^0M_n\cosh\phi+ix^1M_n\sinh \phi}\left|\int dz\int\frac{d\theta}{4\pi} e^{izm\sinh\theta}\frac{1}{\sqrt{m}}\left(\frac{E_n}{\sigma^H}\right)^{\frac{1}{4}} {\rm Ai}\left[(m\sigma^H)^{\frac{1}{3}}\left(\vert z\vert -\frac{E_n}{\sigma^H}\right)\right]\right.,\nonumber\\
&&\times\left.\langle 0\vert \mathcal{A}(x)\vert A,\theta,a_1,b_1;P,-\theta,a_1,b_1\rangle\right|^2+\dots,\label{correlationfunction}
\end{eqnarray}
where we have omitted contributions from higher particle states, which is a good approximation at large distances, $x$. If the operator $\mathcal{A}$ in (\ref{correlationfunction}) is the Noether current $J^L$, we use the two-particle form factor from Eq. (\ref{formfactor}). A similar calculation of correlation functions using intermediate meson states was done for the Ising model in a magnetic field \cite{tsvelik}, and for anisotropic (2+1)-dimensional Yang-Mills theory \cite{twoplusoneform}.

\section{Comments on Numerical Results and Finite Size Effects}
The model we present in this talk has been recently studied numerically by Gongyo and Zwanziger \cite{gongyozwanziger}. The quantities studied in their paper were the expectation value of the Wilson loop, the massive gluon propagator and the order parameter (proportional to the integral over space of the PCSM field, $U$). These objects were evaluated using different lattice spacings, space-time volumes, and values of the coupling constant. From this analysis it was found that at weak Yang-Mills coupling (equivalent to our nonrelativistic limit) the system is in a confined phase, which completely agrees with our results. Their results suggest the existence of a Higgs-like phase as the coupling is increased, which we do not observe. This Higgs-like phase, however, seems to disappear as the volume is increased. We believe this phase to be a finite-volume effect that does not exist at infinite volume.

  We can examine the action (\ref{gaugedpcsm})  in the axial gauge, $A_1=0$ (note that this is different from the completely-fixed axial gauge we previously discussed), and find
\begin{eqnarray}
S=\int d^{2}x \left[\frac{1}{2}{\rm Tr}\,(\partial_{1}A_{0})^{2}+
\frac{1}{2g_{0}^{2}}{\rm Tr}\,(\partial_{0}U^{\dagger}+{\rm i}eU^{\dagger}A_{0})(\partial_{0}U-{\rm i}eA_{0}U)-\frac{1}{2g_{0}^{2}}{\rm Tr}\,\partial_{1}U^{\dagger}\partial_{1}U
 \right].\label{axialgaugeaction}
 \end{eqnarray}
We  now integrate out the gauge field $A_0$  and find
\begin{eqnarray}
S=\int d^{2}x \left(\frac{1}{2g_0^2} {\rm Tr}\,\partial_\mu U^{\dag} \partial^\mu U +\frac{1}{2} \,{j_{0}^{L}}\,\frac{1}{-\partial_{1}^{2}+e^{2}/g_{0}^{2}U^\dag U}
\, {j_{0}^{L}}\right).\label{integrateazero}
\end{eqnarray}
Naively, one would think that since the PCSM field is unitary, $U^\dag U=1$, and the charges are screened. This yields a Higgs-like phase instead of a confined phase. This reasoning is wrong. The reason is that the physical renormalized field is not unitary. The physical field is $\Phi(x)\sim Z(g_0,\Lambda)^{-1/2}U(x)$, where $Z(g_0,\Lambda)$ is a renormalization constant that diverges as $\Lambda\to\infty$ \cite{orland},\cite{asymptotic}. The quantity $\Phi^\dag(x)\Phi(x)$ diverges in the continuum limit and infinite volume. In this sense the completely fixed axial gauge is a much more practical way to see the actual spectrum of the theory. 

There is still a possibility that a Higgs phase could arise at finite volume. We would need to examine the quantity $\lim_{x\to0}\langle 0\vert \Phi^\dag(x)\Phi(0)\vert 0\rangle$ at finite volume. The calculation of finite volume form factors and correlation functions has been discussed in references \cite{finitevolume}. Finite volume effects could regularize this correlation function, and lead to a screened potential between sigma model particles. We would like to point out that the spectrum of the PCSM at finite volume has recently been calculated in Ref. \cite{kazakov} studying its Hirota dynamics.

{\bf Acknowledgements:} I would like like to thank P. Orland for his collaboration and  many discussions throughout this project, and S. Gongyo for explaining some of their work to me during this conference.

\end{document}